# New Cataclysmic Variable 1RXS J073346.0+261933 in Gemini

D. V. Denisenko[1*], A. J. Drake[2], S. G. Djorgovski[2], T. V. Kryachko[3], A. V. Samokhvalov[4], A. Yu. Tkachenko[1]

[1]*Space Research Institute of Russian Academy of Sciences, Moscow*
[2]*California Institute of Technology, Pasadena, USA*
[3]*Astrotel-Caucasus Observatory, Karachay-Cherkessia, Russia*
[4]*Surgut, Russia*

**Abstract**–In course of the search for the optical identifications associated with ROSAT X-ray sources we have found a highly variable object with the very unusual long-term behavior, color indices and high X-ray-to-optical flux ratio. We report the archival photometric light curve from the Catalina Sky Survey, optical spectroscopy from RTT150 and time-resolved photometry from Astrotel-Caucasus telescope. The object appears to be the magnetic cataclysmic variable (polar) with orbital period of P=3.20 hr.

## 1. INTRODUCTION

Twenty years after the ROSAT All-Sky Survey was performed in 1990-1991, many X-ray sources from the bright source catalogue (1RXS) remain poorly studied or unidentified in optical and other bands. Lots of really unusual, highly variable objects of astrophysical interest still remain hidden in it.  Since the catalogue (Voges et al., 1999) was published, dozens of cataclysmic variables were identified among the 18806 sources. See for example, work by Denisenko and Sokolovsky (2011) and references therein.

Cataclysmic variables can be identified from their spectra, which usually show Hydrogen and Helium emission lines, as well as from the light curves (both long term variability on the scales of months and years, and short-term orbital modulations with typical periods of a few hours). In this article we report the results of our analysis of a new remarkable variable object 1RXS J073346.0+261933 on the basis of both photometric and spectroscopic studies.

## 2. IDENTIFICATION

The variability of 1RXS J073346.0+261933 (hereafter referred to as RX J0733.7+2619, or simply J0733+2619) was originally discovered by one of the authors (D.D.) in October, 2007 while routinely checking the DSS (digitized Palomar Observatory Sky Survey) plates of poorly studied bright ROSAT sources. The star USNO-A2.0 1125-05213003 was found within the 12" X-ray error circle (R.A.=07 33 46.28, Decl.= +26 19 26.6, R=18.3, B=18.4) which showed large (~2.5$^m$) variations in brightness between different Palomar plates. Specifically, it was very faint (R~19.5) on the 1997 Jan. 10 POSS-II plate, and bright (R=17.2) on 1999 March 13[th] plate. See Figure 1 for comparison of the two plates mentioned. The X-ray properties of the source (flux 0.081±0.017 cnts/s and the hardness ratio HR1=-1.00±0.10) indicated a possible cataclysmic binary nature. USNO-B1.0 catalog did not provide any measurable proper motion, which if detected would be the final argument against the source being extragalactic.

Following the discovery of variability, Palomar-NEAT (Teegarden et al., 2003) archival CCD images of the object were downloaded from the SkyMorph website

---

[*]*E-mail:* `denis@hea.iki.rssi.ru`

([http://skys.gsfc.nasa.gov/skymorph/skymorph.html](http://skys.gsfc.nasa.gov/skymorph/skymorph.html)). In total, 27 images from nine different nights in the interval 1997 Dec. 26 to 2003 Jan. 24 were used. See Table 1 for the list of measured magnitudes (unfiltered using the R band zero-point). The long-term variability in the range of $17.4^m - 20^m$ was confirmed, with the hints of fast variability on a 15-30 minute scale. Thus, the object was indeed likely to be a cataclysmic variable. However, the NEAT data alone was insufficient to provide an estimate of the orbital period in the system.

Table 1. NEAT photometry of RX J0733.7+2619

| Date, JD | NEAT magnitude | Date, JD | NEAT magnitude |
|---|---|---|---|
| 2450809.01234 | 17.67 | 2452309.77029 | 19.42 |
| 2450809.02226 | 17.45 | 2452319.69204 | 20.35 |
| 2450809.03184 | 17.41 | 2452319.71425 | 19.83 |
| 2452261.79830 | 19.49 | 2452319.73545 | 18.74 |
| 2452261.80061 | 19.04 | 2452327.64956 | 19.07 |
| 2452261.80873 | 19.00 | 2452327.65991 | 18.96 |
| 2452261.81113 | 19.01 | 2452327.67040 | 19.60 |
| 2452261.82157 | 19.13 | 2452337.61903 | 18.98 |
| 2452261.82388 | 19.01 | 2452337.64012 | 19.04 |
| 2452291.93635 | 19.25 | 2452337.66263 | 19.21 |
| 2452291.95732 | 19.06 | 2452662.85890 | 18.98 |
| 2452291.98059 | 18.87 | 2452663.79668 | 19.09 |
| 2452309.72840 | 19.08 | 2452663.87051 | 19.17 |
| 2452309.74944 | 19.02 | | |

3. DATA FROM OTHER CATALOGS

The object was found to be present in 2-Micron All-Sky Survey (2MASS) and Sloan Digital Sky Survey (SDSS) with the following coordinates and magnitudes:

2MASS 07334625+2619260
07 33 46.259 +26 19 26.09, $J$=16.413±0.085, $H$=15.929±0.106, $K_s$=15.706±0.184

The 2MASS *JHK* images were obtained on 1998 Jan. 26 (JD=2450840). This is exactly one month after 1997 Dec. 26 NEAT images (JD=2450809) where the variable was in its bright state. The resulting (*J-K*) index of 0.7±0.2 satisfies a wide range of spectral classes, for main sequence stars this goes from K0V to M5V (Bessell and Brett, 1988).

SDSS J073346.27+261926.2
*u*=20.84±0.06, *g*=20.56±0.02, *r*=19.92±0.02, *i*=18.96±0.01, *z*=18.29±0.02

The SDSS images were taken on 2001 Dec. 18 (JD=2452262, run 2262) when the object was near its minimum light, according to NEAT data (see Table 1). The SDSS color indices are (*u-g*)=0.28, (*g-r*)=0.64, (*r-i*)=0.96 and (*i-z*)=0.67, and are indicative of the binary nature of the system. The rather low (*u-g*) index points to the white dwarf primary, and very large values of (*g-r*) and (*r-i*) correspond to a red secondary component of M3V-M4V subclass.

For comparison, we checked the lists of available color indices of spectroscopically and photometrically confirmed SDSS CVs (Szkody et al., 2004, and references therein) and have found only one close match, SDSS J204827.91+005008.9, with g=19.38, (*u-g*)=0.56, (*g-r*)=0.69, (*r-i*)=1.00, (*i-z*)=0.67, which has orbital period of 4.25 hr. According to Kafka et al. (2010), this is a low-accretion-rate polar with M3/4V secondary, the masses of components being estimated as $M_1 \sim 0.60$ $M_\odot$, $M_2 \sim 0.35$ $M_\odot$.

It is important to note that SDSS J204827.91+005008.9 was also detected by ROSAT All-Sky Survey as an X-ray source 1RXS J204828.2+005022, thus increasing the similarity of two objects. However, J0733+2619 has an X-ray flux about 6 times larger than that of J2048+0050 (8.143E-02 cnts/s vs. 1.368E-02), despite similar optical magnitudes in SDSS catalog.

Among the white-dwarf-main-sequence (WDMS) binaries identified from the spectroscopic SDSS data (Rebassa-Mansergas et al., 2010) there are dozens additional objects with the similar color indices, (*u-g*)>0.5, (*g-r*)>0.6. However, none of these is an X-ray source in 1RXS catalog.

## 4. OPTICAL OBSERVATIONS

The region of RX J0733.7+2619 was observed in December 2008 - January 2009 by T.K. using the private 30-cm Astrotel-Caucasus telescope located at the Mountain Station of Kazan Federal University. About forty 5-min unfiltered exposures were taken during three different nights. Unfortunately the object remained too faint (<19.5$^m$) for us to detect the variability required to measure its period.

Since November 2009 J0733+2619 has been monitored with the 35-cm Bradford Robotic Telescope (http://www.telescope.org/) located at Teide Observatory (Canary Islands, Spain). Images obtained on 2011 Jan. 05 and 13 show the object brightening to 18.0$^m$-17.5$^m$. Using this opportunity, a call for spectral and photometric observations was issued. The optical spectrum with 20-minute exposure and resolution of ~5Å was obtained by A.T. on 2011 Feb. 01 using the 1.5-m Russian-Turkish Telescope at TUG (Turkish National Observatory, Bakirlitepe mountain) and TFOSC imager and spectrometer camera. The calibration and analysis were performed using standard IRAF routines. The spectrum is presented on Figure 2, and shows Balmer emission lines $H_\alpha$, $H_\beta$, $H_\gamma$ and $H_\delta$, as well as a very strong 4686Å He II line.

Fluxes and equivalent widths of the brightest lines in the spectrum of J0733+2619 are listed in Table 2. The error of flux measurements is at 30% level. The He II 4686Å line is the brightest one in the spectrum, its flux being almost 30 percent higher than that of $H_\beta$ line. This remarkable feature was only observed in a few nova-like variables before, for example in SDSS J072910.68+365838.3 (Szkody et al., 2002).

Table 2. Fluxes of the brightest lines in J0733+2619 spectrum. Wavelength is the observed line wavelength as measured by IRAF. Positive EW values are corresponding to emission.

| Wavelength (Å) | Line | Equivalent width (Å) | Flux, $10^{-16}$ erg cm$^{-2}$ s$^{-1}$ |
|---|---|---|---|
| 4100 | $H_\delta$ | 40 | 90 |
| 4340 | $H_\gamma$ | 45 | 98 |
| 4685 | He II 4686Å | 63 | 135 |
| 4861 | $H_\beta$ | 49 | 105 |
| 6563 | $H_\alpha$ | 51 | 90 |

We used the archival data from Catalina Real-time Transient Survey (CRTS, Drake et al., 2009) covering 6 years of observations of J0733+2619 from 2004 to 2010. The light curve of J0733+2619 is updated in real time and available at the following location: http://nesssi.cacr.caltech.edu/catalina/20010730/. Combined CRTS images of the object in a high and a low state are given on Figure 3. Despite showing the significant variability, J0733+2619 was not detected as a transient by the CRTS pipeline, due to its deliberately high triggering threshold for transients.

Having combined Catalina Sky Survey data with NEAT photometry from 2001-2003 (24 points), we have composed the light curve and run period search software to find possible periods. We note that both NEAT and CRTS used unfiltered CCD images, with effectively a wide visible bandpass, and are thus roughly compatible photometrically. Surprisingly, the data suggest an eclipsing-like light curve with a ~940d period (see Figure 4). This finding raised our interest to this object, so we waited for J0733+2619 to become bright again before measuring its orbital period with a moderate sized telescope.

This task was achieved on the nights of 2011 Feb. 8/9, Feb. 11/12 and Apr. 1/2, when the object was observed again by T.K. with Astrotel-Caucasus telescope. In total, seventy-five 300-sec unfiltered exposures were obtained (20, 35 and 20 during the first, second and the third night, respectively). For photometric calibration, we used star USNO-B1.0 1166-0146109 = 2MASS 07340625+2638260 (R.A.=07 34 06.25, Dec.=+26 38 26.1, R1=12.99, R2=13.83, J=12.592, H=12.386, K=12.321) as a comparison star with an adopted magnitude of R=13.41 (average of R1 and R2). The measured magnitudes are given in Table 3. The full (peal-to-peak) amplitude of variability during these observations was measured to be $0.7^m$. A period search was performed by A.S. using *Peranso* (www.peranso.com), and by D.D. using *WinEffect* (by Dr. V. P. Goranskij). Both software packages provided the most likely period to be 0.1334(1)d, or 7.50 cycles per day. However, an alternative period of 0.1395(1)d can not be excluded. The phased light curve folded with the best (0.1334d) period and with a 0.1395d alias is shown in Figure 5.

Table 3. Astrotel-Caucasus photometry of RX J0733.7+2619

| Date, JD | Magnitude | Date, JD | Magnitude | Date, JD | Magnitude |
|---|---|---|---|---|---|
| 2455601.2910 | 17.61 | 2455604.1798 | 17.97 | 2455604.2848 | 17.64 |
| 2455601.2948 | 17.59 | 2455604.1837 | 18.20 | 2455604.2887 | 17.83 |
| 2455601.2987 | 17.55 | 2455604.1876 | 18.23 | 2455604.2965 | 18.19 |
| 2455601.3026 | 17.44 | 2455604.1915 | 18.02 | 2455604.3004 | 18.14 |
| 2455601.3065 | 17.53 | 2455604.1954 | 18.21 | 2455604.3043 | 18.07 |
| 2455601.3104 | 17.46 | 2455604.1993 | 17.99 | 2455653.2837 | 18.18 |
| 2455601.3143 | 17.37 | 2455604.2032 | 18.03 | 2455653.2877 | 18.09 |
| 2455601.3182 | 17.45 | 2455604.2071 | 18.03 | 2455653.2915 | 17.95 |
| 2455601.3258 | 17.59 | 2455604.2110 | 17.87 | 2455653.2954 | 18.08 |
| 2455601.3297 | 17.54 | 2455604.2150 | 17.89 | 2455653.2993 | 18.03 |
| 2455601.3337 | 17.52 | 2455604.2189 | 17.68 | 2455653.3032 | 18.03 |
| 2455601.3376 | 17.58 | 2455604.2228 | 17.83 | 2455653.3071 | 17.96 |
| 2455601.3415 | 17.73 | 2455604.2267 | 17.63 | 2455653.3111 | 17.85 |
| 2455601.3455 | 17.85 | 2455604.2306 | 17.71 | 2455653.3150 | 17.81 |
| 2455601.3493 | 17.82 | 2455604.2346 | 17.67 | 2455653.3189 | 17.82 |
| 2455601.3533 | 17.81 | 2455604.2385 | 17.66 | 2455653.3228 | 17.87 |
| 2455601.3572 | 17.84 | 2455604.2498 | 17.64 | 2455653.3267 | 17.68 |
| 2455601.3611 | 17.90 | 2455604.2537 | 17.64 | 2455653.3307 | 17.78 |
| 2455601.3650 | 18.04 | 2455604.2575 | 17.53 | 2455653.3346 | 17.60 |
| 2455601.3690 | 17.85 | 2455604.2614 | 17.71 | 2455653.3385 | 17.68 |
| 2455604.1604 | 17.97 | 2455604.2653 | 17.84 | 2455653.3424 | 17.68 |
| 2455604.1641 | 18.04 | 2455604.2692 | 17.66 | 2455653.3463 | 17.92 |
| 2455604.1680 | 17.90 | 2455604.2731 | 17.70 | 2455653.3502 | 17.66 |
| 2455604.1719 | 18.16 | 2455604.2770 | 17.83 | 2455653.3541 | 17.78 |
| 2455604.1758 | 18.19 | 2455604.2809 | 17.78 | 2455653.3580 | 17.66 |

## 5. DISCUSSION

Using publicly available data from 4 different catalogs, as well as the archival POSS and NEAT observations, we have discovered a remarkable cataclysmic variable RX J0733.7+2619 showing several types of variability on both short and long timescales. Follow-up observations with three more telescopes allowed us to determine the characteristic periods of this binary system and to obtain its medium-resolution spectrum. The measured value of orbital period 0.1334(1)d, or 3.20 hr, corresponds to a secondary component of M4V spectral class. The data we have gathered so far allow us to tentatively classify the system as the magnetic cataclysmic variable (polar), probably a low accretion rate polar. We will continue to monitor this system to better study its behavior.

As mentioned above, the CRTS and NEAT data provided us with an eclipse-like light curve with a ~940d period. The nature of this nearly periodic fading remains a mystery so far and requires further study. It should be noted that the minima of the light curve are very narrow, and better described by $|\sin(t/P)|$ function than a plain harmonic wave. Such a behavior (round maxima and sharp minima) could be explained by a changing viewing geometry of the system, for example, due to precession of the accretion disc. When the system is viewed edge-on, the rate of change of the inclination angle is at its maximum value, thus the "eclipses" are short. Other possible models such as a) semi-regular pulsations of the secondary, b) eclipses by the third body, or c) periodic changes of the accretion rate are less attractive, however, they cannot be completely ruled out. For example, similar long-term behavior of FS Aur can be explained by the close binary orbit modulation by the presence of a third body on a circular orbit (Tovmassian et al., 2010). A long term monitoring campaign covering the spectra of J0733+2619 at the various phases of ~940d period is encouraged.

## ACKNOWLEDGEMENTS

Authors thank Kadir Uluc (TUG) for obtaining the spectrum of J0733+2619. DD and AT would like to thank the Program of support of the leading scientific schools (grant NSh-5069.2010.2). The CRTS survey is supported by the U.S. National science foundation grant AST-0909182. AJD and SGD wish to thank their CRTS collaborators for sustained efforts in conducting the survey. Authors also would like to thank the anonymous referee for his useful comments.

FIGURES

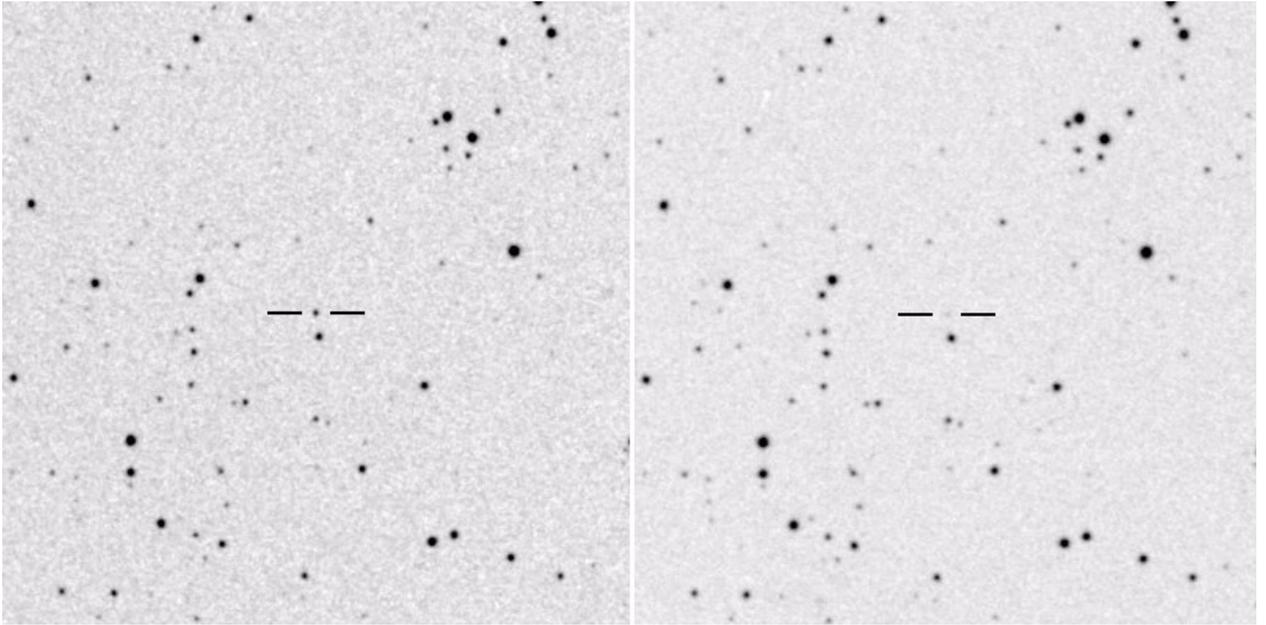

Figure 1. 400″x400″ finder chart of J0733+2619. Left: Red Palomar plate taken in 1999. Right: Red Palomar plate of 1997. North is up, East is to the left.

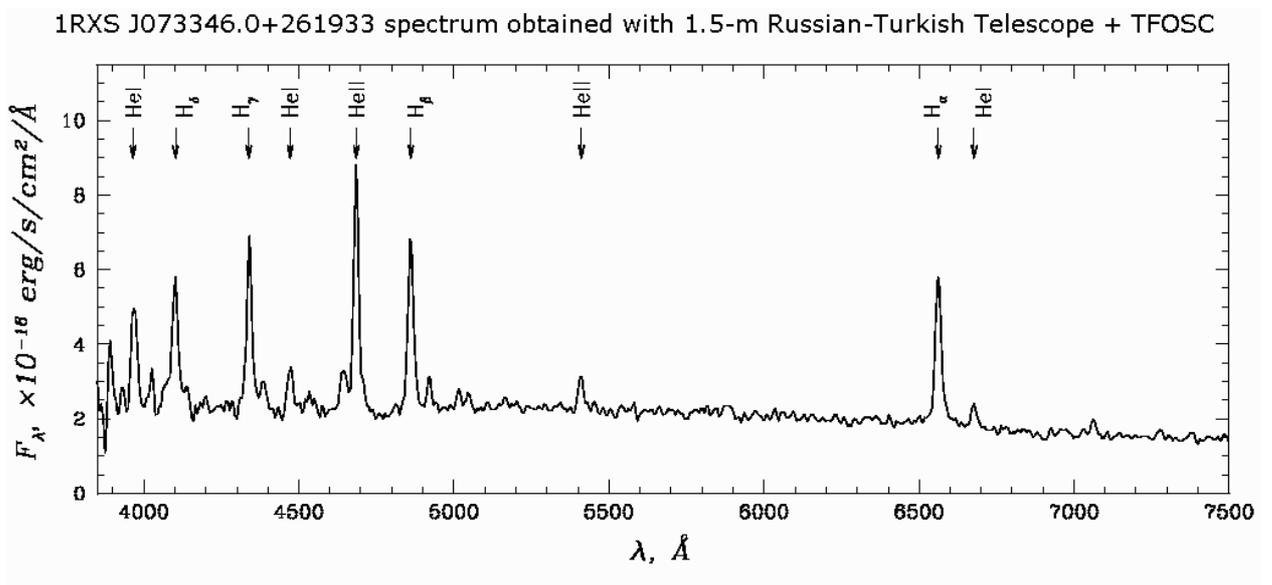

Figure 2. RTT150 spectrum of J0733+2619 obtained on 2011 Feb. 01

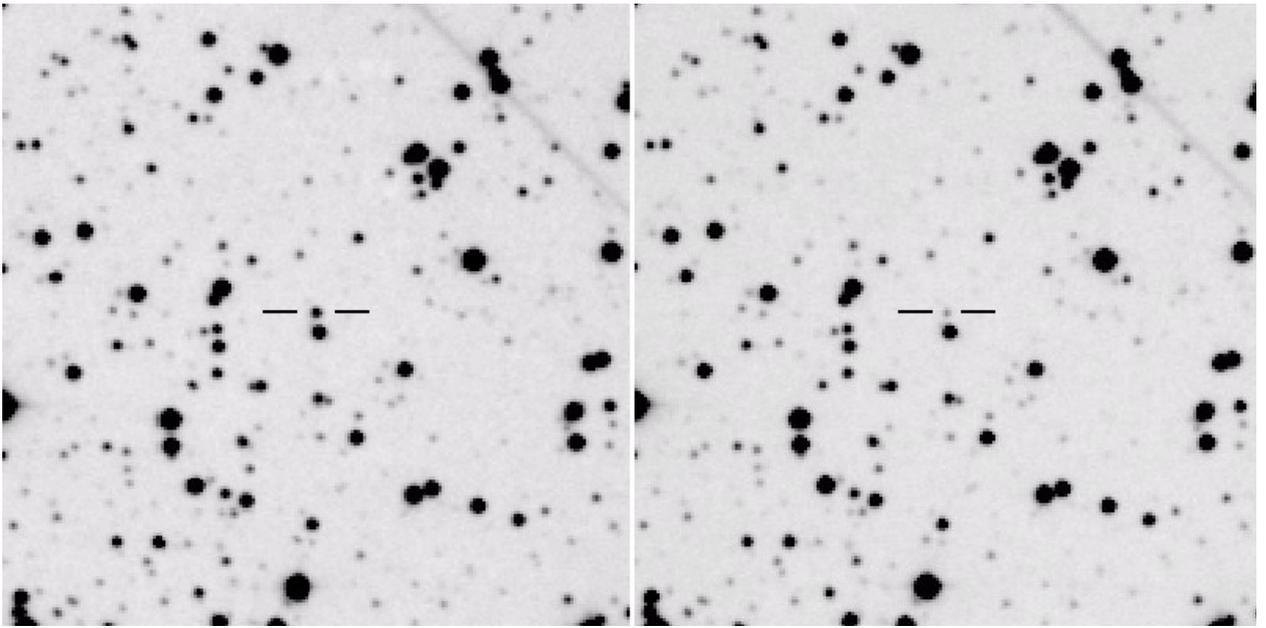

Figure 3. CRTS images of J0733+2619 in a high state (left) and in a low state (right). Field of view is 500″x500″, North up, East left.

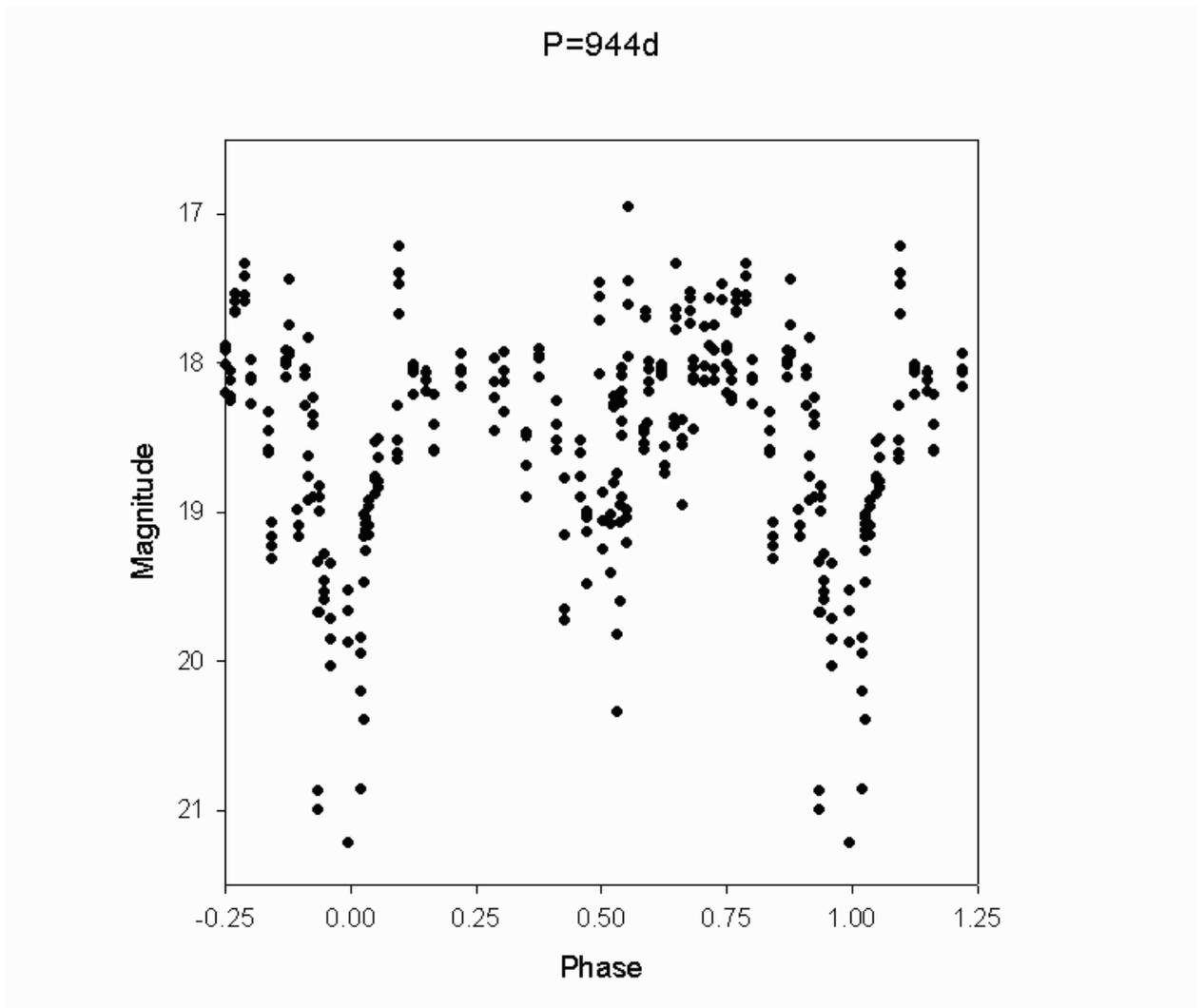

Figure 4. Archival NEAT+CRTS light curve of J0733+2619 folded with the best period

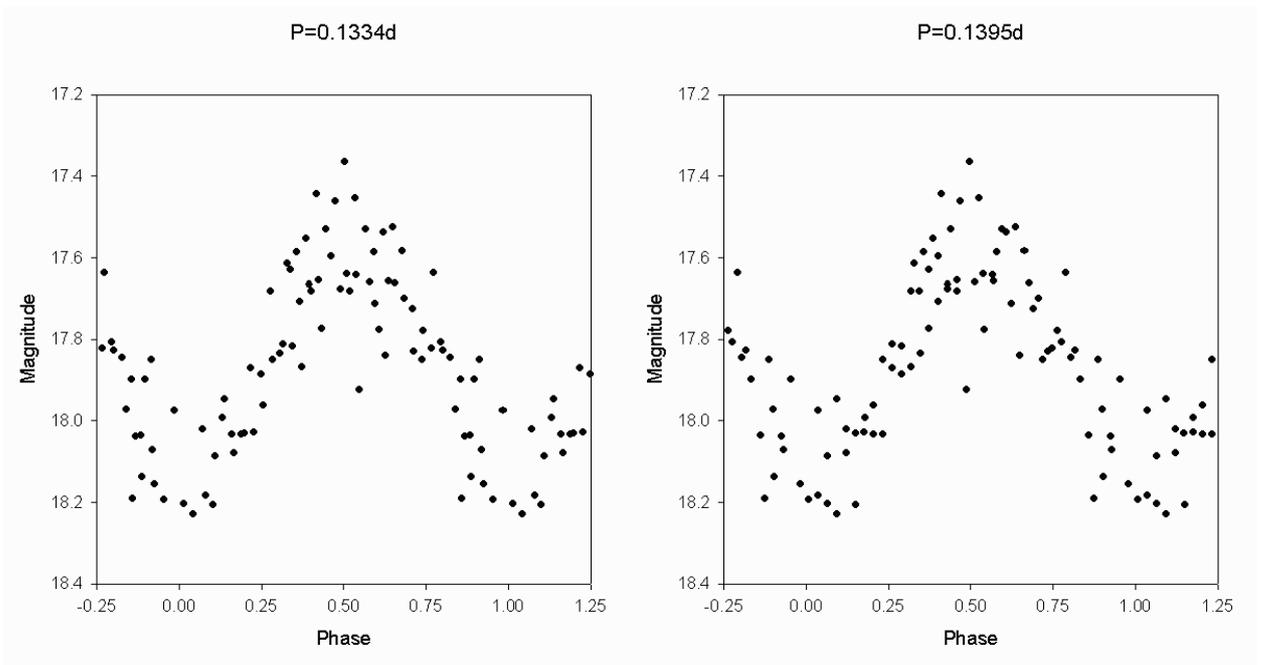

Figure 5. Astrotel-Caucasus light curve of J0733+2619 folded with the best orbital period P=0.1334d (left) and with a possible alias P=0.1395d (right)